\documentclass{article}
\usepackage{emulateapj,graphics,onecolfloat,amsmath}
\usepackage{epsfig}
\usepackage{subfigure}

\input epsf
\lefthead{Schulz et. al.}
\righthead{Hybrid Plasma Modeling}

\begin{document}
\twocolumn[
\title {Hybrid Particle-Fluid Modeling of Plasmas}
\author {A.E. Schulz$^{1,2}$, A.D. Greenwood$^{1}$, K.L. Cartwright$^{1}$, 
P.J. Mardahl$^{1}$}
\affil{$^{1}$Air Force Research Laboratory, Directed Energy Directorate,
Kirtland AFB, NM 87117}
\affil{$^{2}$Department of Physics, Harvard University,
Cambridge, MA 02138}

\begin{abstract}
\noindent
There are many interesting physical processes which involve the 
generation of high density plasmas in large volumes. 
However, when modeling these systems numerically, the 
large densities and volumes present
a significant computational challenge.   One technique
for modeling plasma physics, 
the particle in cell (PIC) approach, is very accurate 
but requires increasing computation 
time and numerical resolution as the density of the plasma grows. 
In this paper we present a new 
technique for mitigating the extreme computational load as the 
plasma density grows by combining 
existing PIC methods with a dielectric fluid approach.
By using both descriptions in a hybrid 
particle-fluid model, we now can probe 
the physics in large volume, high density regions. 
The hybrid method also
provides a smooth transition as the plasma density increases and 
the ionization fraction grows to 
values that are well described by the fluid description alone.  
We present the hybrid technique and demonstrate the validity of the physical 
model by benchmarking against a simple example with an analytic solution.    
\end{abstract}

]

\rightskip=0pt
\section{Introduction}

Particle in cell (PIC) methods enjoy great success in modeling devices that 
include moderately dense plasmas.  However, as the plasma density becomes
high in a large volume, the number of particles to track becomes 
computationally prohibitive.  Reducing the number of particles by creating 
larger ``macro particles'' introduces unacceptable numerical error.  
Alternatively, high density plasmas in large volumes can be modeled using
a dielectric fluid description.  
This requires integrating over the time dependent
distribution function of
electrons in the plasma, which for computational simplicity is often assumed to 
be well approximated by a Maxwellian.  
However, this is not accurate 
for characterizing the formation of very dense plasmas, where often times 
the distribution function is not entirely known.  For example,
some physical processes such as 
air breakdown phenomena involve energy distributions
that are partially Maxwellian but include a long, high energy tail.  The 
particles in the high energy tail are those responsible for the majority of 
interactions that lead to a qualitative change in physical behavior
\cite{Nicholson}.  Thus,
a fluid description that neglects the high energy tail fails to capture the 
physics of interest.  We propose a hybrid plasma description that 
simultaneously employs the fluid and PIC treatments which 
has the potential to capture
the relevant physics in a tractable computational time.  

In addition to reducing the number of particles that must be tracked by 
the PIC treatment, the particle-fluid hybrid approach also allows 
computations to be performed on a much coarser grid while preserving the 
physics.  In general, explicit PIC computations become inaccurate when the 
resolution of the grid becomes comparable to the Debye length of the plasma. 
In the hybrid scenario, however, the size of the grid need only be 
comparable to the Debye length of the partial-plasma comprised of the 
particles in the high energy tail.  Since the Debye length varies 
inversely as the $\sqrt{{\rm density}}$, and the 
high energy tail contains only a 
small fraction of the total density, the grid resolution can be significantly
reduced.  The grid resolution requirements are determined by the 
Debye length of the partial-plasma and by the length scale required 
to resolve the spatial gradients in the dielectric fluid treatment.
The division criterion used to separate the plasma into particles and 
fluid will determine which of these conditions will require the
greater resolution.

In the hybrid description we use a fluid model for particles that fall 
within a Maxwellian energy distribution, and PIC for particles in the high
energy tail.
We have added 
a dielectric fluid description of a plasma to the Improved 
Concurrent Electromagnetic Particle In Cell (ICEPIC) code (\cite{Birdsall};
\cite{Luginsland}).  To 
ensure physical accuracy the PIC and 
fluid descriptions are tested independently. 
A test problem with a simple analytic solution is used to benchmark
the performance of the particle, dielectric
fluid, and particle-fluid hybrid treatments.  
All three approaches are shown to properly reproduce the correct dispersion
relation when electromagnetic plane waves are launched through a 2-D box 
containing hot or cold plasma.

\section{Method}
\subsection{The Particle In Cell (PIC) treatment}

ICEPIC computes the time advance of the 
magnetic field according to Faraday's law, and the electric field according 
to Ampere-Maxwell's law.  The discreet form of these equations used in 
ICEPIC are designed to preserve the constraint 
equations $\nabla \cdot {\bf B} = 0$
and $\nabla \cdot {\bf E} = \rho/ \epsilon_0$ as long as the initial data 
satisfies these constraints. 
The particles used in ICEPIC are ``macro-particles''
that represent many charged particles (electrons and/or ions) with a 
position vector ${\bf x}$ and a velocity vector ${\bf v}=d{\bf x}/dt$.  The 
relativistic form of Lorentz's force equation is used to determine the 
particle's velocity:
\begin{equation}
{\bf F} = m{d \gamma {\bf v} \over dt} = q\left({\bf E} +{{\bf v} \over c} 
\times {\bf B} \right)
\end{equation}
where $\gamma$ is the the usual relativistic factor of $(1-v^2/c^2)^{-1/2}$, 
and $q$ and $m$ are the charge and mass of the particle. 

ICEPIC uses a fixed, Cartesian,
logical grid to difference the electric and magnetic 
field equations. 
The vector quantities ${\bf E}$, ${\bf B}$, and ${\bf J}$
are staggered in their grid location using the technique of \cite{Yee}.
${\bf E}$ and ${\bf J}$ are located on the edges of the primary grid, whereas
${\bf B}$ is located on the faces of the primary grid.  An explicit leap-frog
time step technique is used to advance the electric and magnetic fields 
forward in time.  The advantages of the leap-frog method are simplicity and 
second-order accuracy.  The electric field advances on whole integer time 
steps whereas the magnetic field and the current density advance on half 
integer time steps.  

The three components of the momentum and position of each particle are updated
via Eq. (1) using the Boris relativistic particle push algorithm 
\cite{Boris}. The particle equations for velocity and position are also 
advanced with a leap-frog technique.  The velocity components are advanced
on half integer time steps, and the particle positions are 
updated on integer time steps.  The current density weighting employs an 
exact charge conserving current weighting algorithm by \cite{Villa}, enforcing
$\nabla \cdot {\bf E} = \rho / \epsilon_0$.  Once the particles' positions 
and velocities are updated and the new current density is updated on the grid, 
the solution process starts over again by solving the field equations. 
 
\subsection{The Dielectric Fluid Model}

For very dense plasmas, it is often a very good approximation to treat 
the plasma as a dielectric fluid.  In the presence of EM 
radiation, this approximation is good for time-scales over which the 
fluid moves a negligible amount.  The dielectric constant used in the 
fluid approximation is (in terms of the permittivity of free space 
$\epsilon_0$, the collision frequency $\nu_s$, and the frequency of the 
electromagnetic radiation $\omega_{\text{\tiny EM}}$)
\begin{equation}
\epsilon=\epsilon_0 \left( 1-{\omega_p^2 \over \omega_{\text{\tiny EM}} 
(\omega_{\text{\tiny EM}}
+ i \nu_s )} \right)
\end{equation}
where the plasma frequency $\omega_p$ depends on the density $n_0$ 
and species ($m$, $q$) of the ionized particles, and is given by 
$\omega_p^2=n_0 q^2 / m \epsilon_0$.

To model high density plasma in the time domain, consider Ampere's law with 
the plasma dielectric constant;

\begin{align}
\nabla \times& {\bf B}({\bf x},\omega_{\text{\tiny EM}}) = \nonumber
\\ &\mu_0 \epsilon_0 
\left( -i \omega_{\text{\tiny EM}} +
{\omega_p^2 \over \nu_s - i \omega_{\text{\tiny EM}}} \right) {\bf E}({\bf x},\omega_{EM})
\end{align}

The equivalent expression in the time
domain is given by

\begin{align}
\nabla\times&{\bf B}({\bf x},t)= &\nonumber \\
&\mu_0 \epsilon_0 {\partial {\bf E}(
{\bf x},t) \over \partial t}+\omega_p^2 \mu_0 \epsilon_0 \int_{-
\infty}^t e^{- \nu_s (t-\tau)}{\bf E}({\bf x},\tau) d\tau
\end{align}

Differentiating Eq. (4) with respect to 
time and substituting the result back into Eq. (4) eliminates 
the convolution integral and yields the 
following expression used to define the field update equations in the 
dielectric fluid model.

\begin{align}
{\partial^2 {\bf E}({\bf x},t) \over \partial t^2}&+\nu_s {\partial {\bf E}(
{\bf x},t) \over \partial t}
+\omega_p^2 {\bf E}({\bf x},t) \nonumber \\
&={1 \over \mu_0 \epsilon_0} 
\nabla \times \left( {\partial {\bf B}({\bf x},t) \over \partial t} + \nu_s
{\bf B}({\bf x},t) \right)
\end{align}

Our implementation of this equation, together with 
Faraday's law used for updating ${\bf B}$
in our treatment, is displayed in the inset on
the following page.  It can be shown that this method exhibits 2nd order
accuracy, which is demonstrated later in this paper.  We have determined that this update is stable if the condition (\ref{eq:condition}) is satisfied:
\begin{align}
\cos(\omega_p\Delta_t) \le (X-1)/(X+1) \label{eq:condition}
\end{align}
where
\begin{align}
X = \frac{4c^2}{\omega^2} \left (\frac{1}{\Delta_x^2} + \frac{1}{\Delta_y^2} +\frac{1}{\Delta_y^2}\right). \nonumber 
\end{align}
\begin{center}
\begin{table*}[th]
\begin{tabular*}{7 in}{|l|}
\hline
\\
The equations used in ICEPIC to perform the ${\bf E}$ and 
${\bf B}$ field updates. \\ \\
${\bf E}^{n+1} = 2 f_1(\Delta_t,\omega_p,\nu_s){\bf E}^n - 
e^{- \nu_s \Delta_t}
{\bf E}^{n-1} + {1 \over \mu_0 \epsilon_0 \omega_p^2 \Delta_t} \nabla \times
\left[ \nu_s \Delta_t \left( {\bf B}^{n+1/2} + {\bf B}^{n-1/2}\right)
f_2(\Delta_t, \omega_p, \nu_s) \right.$ \\ 
\hspace{1.95 in} $ \left. +2\left({\bf B}^{n+1/2} - 
{\bf B}^{n-1/2}\right) \left[ \left(1-{\nu_s^2 
\over \omega_p^2} \right) f_2(\Delta_t, \omega_p, \nu_s) +
\nu_s \Delta_t e^{- \nu_s \Delta_t /2} \;{\rm sinh} \left(\nu_s \Delta_t /
2\right) \right] \right]$
\\
\\
${\bf B}^{n+1/2} = {\bf B}^{n-1/2} - \Delta_t \nabla \times {\bf E}^n$
\\
\\
where 
\\
\\
$f_1(\Delta_t, \omega_p, \nu_s)=e^{- \nu_s \Delta_t/2}\; {\rm cosh} \left( 
{\Delta_t \over 2}\sqrt{\nu_s^2 -4\omega_p^2} \right) $
\\ 
\\
$f_2(\Delta_t, \omega_p, \nu_s)=e^{- \nu_s \Delta_t/2} \left( {\rm cosh}
\left( \nu_s \Delta_t /2 \right) 
- {\rm cosh} \left( {\Delta_t \over 2}
\sqrt{\nu_s^2 -4\omega_p^2} \right) \right)$  
\\
\hline
\end{tabular*}
\end{table*}
\end{center}

\subsection{Numerical Methods to Create Hybrid Models}

Developing a consistent way to divide the simulated plasma into fluid and 
particle portions is one of the more complicated aspects of this approach. 
Several details need to be considered.  Most importantly, the 
distinction between the fluid and particle descriptions is not a physical
one, but rather a computational necessity.  As such, it is crucial that 
dividing the plasma into two separate populations does not introduce
any spurious observable behaviors in the physics.  The real-world plasma
is not divided, so the particles and the fluid must be compelled to 
interact as a single species.  This manifests itself in two ways; first, 
the mechanism for exchanging particles into fluid (and vice versa) must 
be seamless enough that it does not affect the global properties of the 
plasma. Second, the fluctuations in density, 
pressure, and temperature in the fluid 
must affect the dynamics of the PIC particles in exactly the same way 
as if those fluctuations had occurred in a purely PIC model. 
In the special case where the 
density contained in the fluid is significantly greater than the 
density in the particles, it may be a good approximation to neglect these
fluctuations in the PIC particles.  This is discussed later in section 
2.5.  

Another important priority in the development of a useful hybrid model 
is to determine the optimal way to divide the plasma.  An 
appropriate criterion must be found for the exchange of PIC particles with 
fluid density.  It is important to treat as much of the plasma as possible 
with the fluid model, since this minimizes computation time.  On the 
other hand, if the decision criterion is computationally expensive, it would
not be sensible to perform the particle-fluid exchange in every time step.  
A balance must be found between the time saved by moving particles into the 
fluid and the time spent deciding whether the particles can be moved 
without sabotaging the accuracy of the simulation. 

Finally, although collision physics have been added to the fluid and PIC 
models independently, collisions between particles and elements of fluid 
have not yet been modeled and tested for the current simulations, and are 
a subject for future study.  Ultimately, 
these will affect the balance of the energy 
and the density in the fluid. More extensive future modeling will 
include the effects of mobility, diffusion, ionization with the background 
gas, and feedback heating from the external fields.   

\subsection{Discussion of Hybrid Errors and Limitations}

Quantitatively, implementing a hybrid approach is accomplished by 
dividing the distribution of the plasma $f$
into two separate populations, $f_1$
and $f_2$.   The distributions $f_1$ and $f_2$ sum to the total 
distribution $f=f_1+f_2$ and for simplicity are positive definite such that 
$f_1<f$ and $f_2<f$.
From a fluid perspective, the continuity equation and fluid 
force equations are obtained by taking the first two 
moments of the Vlasov equation, 
which dictates the evolution of the plasma distribution function in a 
6+1 dimensional (${\bf x}$,${\bf v}$,$t$) phase space. 
\begin{eqnarray}
\partial_t f({\bf x},{\bf v},t) + {\bf v} \cdot \nabla_x f + 
{q \over m} \left( {\bf E} + {{\bf v} \over c} \times {\bf B} \right)  
\cdot \nabla_v f = 0
\end{eqnarray}
Here
the effects of collisions are ignored.  Since each term in the Vlasov 
equation is linear in $f$, the continuity and force equations 
contain no terms proportional to $f_1f_2$, and can be satisfied 
by evolving the two populations separately.  For the test problem presented in
the following section, one of the two populations is evolved using the PIC 
technique, while the other population is treated as a dielectric fluid.  The 
PIC technique automatically satisfies the continuity and force equations, 
but our implementation of the dielectric fluid does not contain two of the 
terms in the force equation; the $(q/mc)n({\bf v} \times {\bf B})$ term and 
the $\nabla_x \cdot (n \langle {\bf v v} \rangle)$ term.  The justification
for neglecting these terms 
is that drift 
velocities in the fluid will be small, and components of the plasma 
with high energies and velocities will be modeled with PIC particles.  

When collisions are added the description becomes more complex because 
collisions will have to be modeled between populations, as well as 
within each population.  In this case linearity is lost, and it may not 
be possible to 
decouple the evolution of the two populations.  
It may, however, be acceptable to 
make certain simplifying assumptions, such as $n_{{\rm particles}}<<
n_{{\rm fluid}}$.  In this limit, the contribution to the fluid dynamics 
from fluid-fluid collisions far exceed 
the contribution from 
fluid-particle collisions, and the latter might safely be neglected.  
The error introduced by this assumption will depend on the relative 
distributions $f_1$ and $f_2$.   When implementing the model of collisions
it is necessary to 
estimate analytically the error introduced as a function of the 
particle density $n_{{\rm particles}}$, and re-distribute the populations
when a pre-determined threshold for error is reached.  The error analysis
of this method to 
incorporate collisions into the hybrid plasma model is currently under 
investigation.    

\subsection{The Dispersion Relation}

To demonstrate that the numerical methods discussed above are valid
treatments of high density plasma physics in large volumes, we have employed
these methods to calculate the dispersion relation of electromagnetic plane
waves traveling through a plasma in a two dimensional box.  This is simple 
enough that an analytic expression for the dispersion relation exists, and 
also small enough to make an explicit PIC treatment practical, even for 
very high densities.  In this way we are able to directly compare the 
theoretical prediction of the dispersion relation to the 
results of the individual particle (PIC) and dielectric fluid treatments, and
to the PIC-fluid hybrid computation.  
We perform all the calculations (analytic and numerical) in the 
limit of non-relativistic thermal velocities in the plasma, a good 
approximation in clouds of plasma recently ionized by directed energy sources. 
 
The analytic expression relating the frequency $\omega_{\text{\tiny EM}}$ 
of transverse electromagnetic 
waves to the wave number $k$ in a cold plasma is given by the following 
dispersion relation (see e.g. \cite{Chen}):
\begin{equation}
\omega_{\text{\tiny EM}}^2=\omega_p^2+k^2c^2 \;\;\; .
\end{equation}
Here the plasma frequency $\omega_p$ depends on the density and 
and species of the plasma and is given in section 2.2. 
and the collision frequency $\nu_s$ is assumed to be zero. 
The expression for the dispersion relation becomes significantly more 
complicated for a warm plasma, because thermal velocities allow for 
a coupling between adjacent regions in the equations of motion, due to 
local pressure gradients.  However, in the non-relativistic limits being 
considered here, the change in the dispersion relation due to thermal 
fluctuations is less than 0.01\%, and we neglect the effect.   

We have performed simulations in a simple geometry in order to extract
the numerical dispersion relation and compare the results to the theoretically
expected curve.  A row of dipole antennas 
polarized in the $\hat{z}$ direction are lined up at the left end of a 
long box filled with a relatively dense cold plasma ($1.43 \times 10^{17}
\text{electrons} / m^3$ corresponding to a plasma frequency of $\omega_p=21.1 
\times 10^9$ rad/sec).  The antennas are 
driven in phase with an oscillating current, at a peak 
amplitude of $0.1$ Amps,
to generate a plane wave in the plasma. The top and bottom edges of the box
are defined by metal boundaries. There is a perfectly matched layer (PML)
on both the left and right sides of the box
to prevent any reflection of incoming EM waves, simulating an infinite 
domain in that direction. 
A cartoon of the geometry is shown in Figure (1). 
Simulations run from one to four light crossing 
times.  At one light crossing time there are observable
transient effects while the current in the antennas ramps up, while at 
4 light crossing times there is significant beating from reflections
off of the interior of the plasma.  
While these effects make it difficult to quantify properties of the 
plasma such as the skin depth or transmission and reflection 
coefficients, neither of them affected the 
calculated values of $k$ 
associated with the frequency of the incoming plane wave. 

\begin{figure}[h]
\centerline{\epsfysize=3.5cm \epsfbox{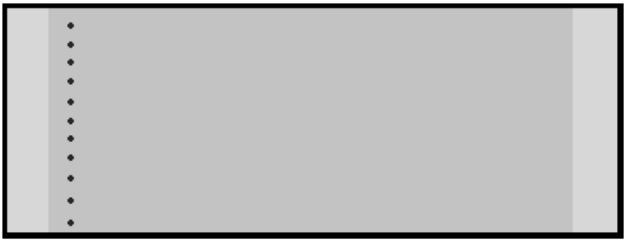}}
\caption{A 2-D box with metal on top and bottom, PML on left and right, and 
a cold plasma in the interior.  The line of dipole antennas (polarized in 
$\hat{z}$) generating the 
EM plane wave is marked with dots on the left side.  The dimensions of the 
box are $1.4 m \times 0.5 m$}
\label{fig:geom}
\end{figure}

To bench-mark the performance of the PIC-fluid hybrid treatment, we first
run simulations in this geometry with the plasma consisting of only 
PIC particles, then with only dielectric fluid.  We then test a model where 
half of the plasma is treated as PIC particles and the other half modeled 
as a fluid.  This division is in each cell, rather than in different 
spatial regions in the box.  There is no mechanism for exchange between 
fluid and particles.  The simulations are run using many 
frequencies of the EM plane waves, thus providing several different
data points to fit to the theoretical dispersion curve.  To calculate the
value of $k$, the pixels are averaged in the $\hat{y}$ direction, and a
spatial Fourier transform taken of the resulting 1-D ($\hat{x}$) array.  The
error in the resulting wave number $k$ is inherited from the finite box size.

\vspace{1 cm}
\section{Results}

The dispersion relations obtained from the explicit PIC, dielectric fluid,
and fluid-PIC hybrid models are displayed in Figure (2) along with the 
theoretical value from Eq. (7).  
Although the value of k is the dependent variable in our analysis, we 
plot the wavenumber on the x axis so the dispersion relation 
takes its familiar form.  All three modeling techniques 
yield the same values of k within the resolution of the simulations; they 
agree to within $\pm 2.5$ cycles/$m$,
approximately the width of the symbols used 
in the plot.  
\begin{figure}[h]
\centerline{\epsfysize=7cm \epsfbox{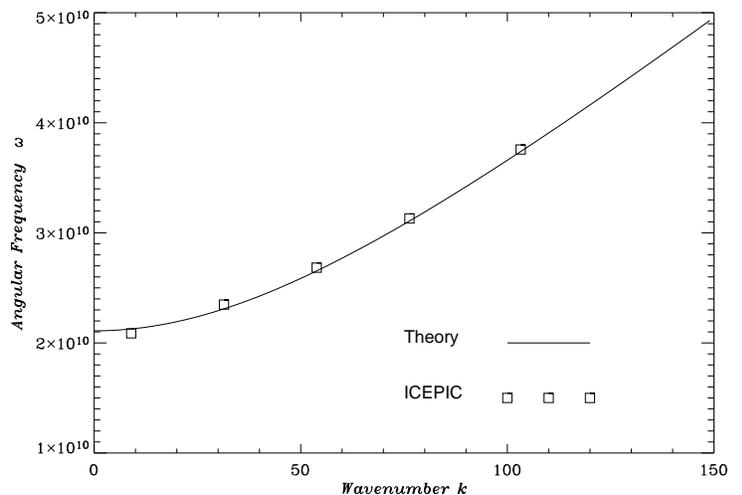}}
\caption{The dispersion relation for E-M plane waves traveling through
a cold plasma in a 2-D box.  This plot shows simultaneously the results
for the PIC, dielectric fluid, and PIC-fluid hybrid models.  All three 
models yielded results that fell in the same bin in k 
with an error bar of 
$\pm 2.5$ cycles/$m$, and are labeled ICEPIC in the plot. }
\label{fig:disp}
\end{figure}

The transmission of electromagnetic radiation through 
a plasma as a function of the plasma density and chemical properties  
is another interesting quantity to consider when evaluating the 
validity of the particle-fluid hybrid model.  We are particularly 
interested in probing the regime where the frequency of the 
electromagnetic radiation is comparable to the plasma frequency.  
To calculate the transmission coefficient as a function of 
the frequency of the plane wave, we modified the geometry
of the previous
simulation to include three regions; a vacuum region containing
the antennas, a region in the center containing a cold plasma with the 
same density, and a third vacuum region on the right. The amplitudes are 
calculated by taking the Fourier transform of 
the time history of the quantity $({\bf E}-c{\bf B})
/2$, which represents the magnitude of the right-going electromagnetic
wave in a vacuum. The ratio of the 
amplitudes in the two vacuum regions is plotted as a function of 
frequency in Figure (3).  
\begin{figure}[h]
\centerline{\epsfysize=6.5cm \epsfbox{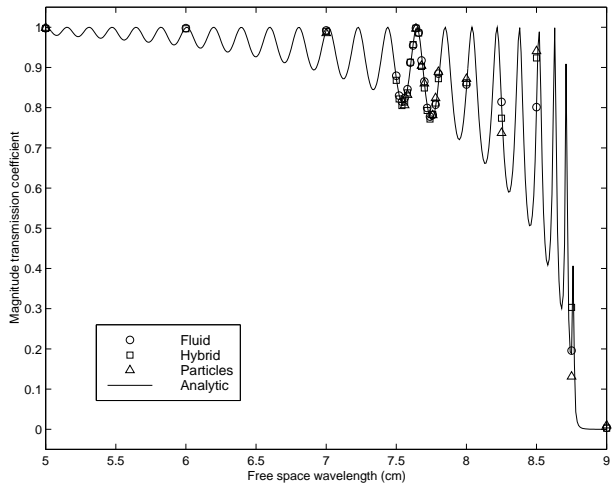}}
\caption{The fraction of a plane wave transmitted through a $7 cm$ region
of plasma of density $1.43 \times 10^{17}$ electrons/$m^3$.}
\label{fig:geom}
\end{figure}
The theoretical curve in Figure (3) is obtained from the boundary 
matching at the two vacuum-plasma interfaces.  The positions of the 
maxima and minima in the transmission coefficient depend upon 
the width of the plasma region, in this simulation $7 cm$.  

Although there is striking agreement between the computed values of the 
transmission coefficients and dispersion relation with the theoretically
predicted curves, a formal demonstration of the convergence of the code is
needed to be sure that the resolution is sufficient to accurately model the
physics.  We have examined the transmission at three different wavelengths
of incident radiation, corresponding to a local maximum ($\lambda=7.0$ cm), 
a local minimum ($\lambda=7.73$ cm), and an area of high slope nearer to the 
plasma frequency($\lambda=8.25$ cm).  For each of these cases, the 
spatial resolution of the grid was varied from 6 to 40 cells per free-space
wavelength of the incident radiation.  When there were particles in the 
simulations, the total number of particles was kept fixed.  At the finest 
spatial 
resolution there were enough particles per cell to maintain numerical 
accuracy.  The results of this investigation are summarized in Figure (4),
which demonstrates 2nd order convergence to the theoretical value. 
From Figure (4) we conclude that at a resolution of 20 cells per wavelength, 
the resolution used when generating Figures (2)and (3), the code has 
converged to produce less than 6\% worst-case error from the theoretical value. 
Seven of nine runs at this resolution produced errors less than 1\%.
Simulations yielding errors larger than 1\% correspond to a wavelength of 8.25 cm, which is in a high slope region of the theoretical curve.  Here, the small errors in plasma width as represented by the grid lead to larger errors in the simulated transmission.
The
reader should be reminded that differences between the computed and theoretical
values have contributions from error in the modeling, approximations in
the calculation of the theoretical value, and numerical errors due to finite
resolution.   
\begin{figure}[t]
\centering
\subfigure[$\lambda_{\rm vac} = 7 cm$]{\epsfig{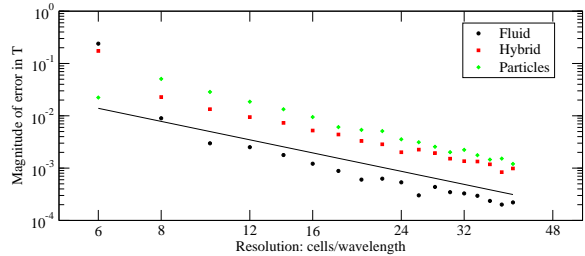}}
\subfigure[$\lambda_{\rm vac} = 7.73 cm$]{\epsfig{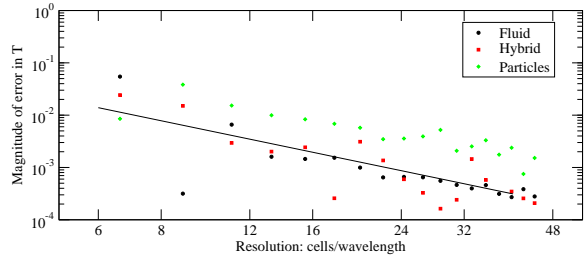}}
\subfigure[$\lambda_{\rm vac} = 8.25 cm$]{\epsfig{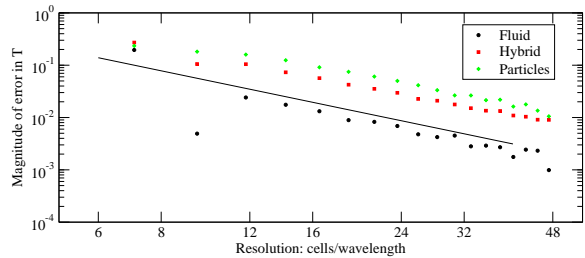}}
\caption{Error in the transmission coefficient, T, as a function of 
grid resolution for fluid, particle, and particle-fluid hybrid models, 
evaluated at three different incident wavelengths.  The solid lines show 
the slope of 2nd order convergence.}
\label{godfreyfilt}
\end{figure}

\vspace{1 cm}

\section{Conclusions}

We demonstrate the validity of a powerful new technique that
allows numerical simulations of high density plasmas in large volumes to 
be computationally tractable in reasonable times.  Among other 
applications such as tokamaks, plasma processing, atmospheric plasmas 
(lightning, red sprites, blue jets), plasma-display plasmas, and lighting,  
this approach will 
aid in the exploration of many unanswered questions about 
Radio Frequency (RF) breakdown
in air.
We are particularly interested in understanding the mechanisms of breakdown.
High density plasmas in large volumes occur relatively frequently when
studying High Power Microwave (HPM) or other directed energy devices. 
Sources of high power microwaves can cause RF breakdown in 
air or other gaseous media in the vicinity of the antenna.  It is
extremely useful for design exploration if such breakdown processes could 
be modeled computationally.  
Several details can be added to make this tool even more useful.  Collisional
interactions can be implemented in both the PIC and fluid treatments, taking
care that collisional interactions between the populations are accurately 
modeled.
The treatment should also be adapted to accommodate relativistic particle 
velocities.  Reasonable mechanisms need to be added for interchange of 
particles to fluid, and vice versa.  These mechanisms must be treated with 
care so as not to introduce any spurious effects into the physics. 

When collisions and gas chemistry 
have been fully implemented we will use this method to
explore the details of the plasma formation 
as a function of the pulse width, the 
${\bf E}$ field intensities causing the breakdown, and the properties 
of the background gas (density, impurities).  We hope to quantitatively 
determine when the breakdown occurs and what fraction of the RF pulse is 
reflected, an effect that causes tail erosion.
By determining the ionization states 
that are 
produced, and the densities that are reached, we will be able
to describe how the global properties of the medium will change.  

\bigskip
The authors would like to thank Matthew Bettencourt, 
Peter J. Turchi and Kyle Hendricks for
useful discussions on this work.
This research was supported in part by Air Force Office of Scientific 
Research (AFOSR).


\begin{thebibliography}{99}

\bibitem[Birdsall \& Langdon 1985]{Birdsall}
Birdsall, C.K., Langdon, A.B., {\it Plasma Physics Via Computer Simulation},
McGraw-Hill, New York, (1985)
\bibitem[(Boris 1970)]{Boris}
Boris, J.P., ``Relativistic Plasma Simulation-Optimization of a Hybrid Code,''
{\it Num. Sim. Plasmas}, Navan Res. Lab., Wash D.C., (1970) pp3-67
\bibitem[Chen 1984]{Chen}
Chen, F.F., {\it Introduction to Plasma Physics and Controlled Fusion Volume 1},
Plenum Press, New York, (1984)
\bibitem[Luginsland \& Peterkin 2002]{Luginsland}
Luginsland, J.W., Peterkin, R.E., ``A Virtual Prototyping Environment for 
Directed Energy Concepts,'' {\it Computing in Science \& Engineering}; 
March-April
(2002); vol.4, no.2, pp.42-9
\bibitem[(Nicholson 1983)]{Nicholson}
Nicholson, Dwight R., {\it Introduction to Plasma Theory}, Krieger Publishing
Company, (1983)
\bibitem[Villasenor \& Buneman (1992)]{Villa}
Villasenor, J., Buneman, O., ``Rigorous Charge Conservation for Local 
Electromagnetic Field Solvers,'' {\it Comp. Phys. Comm}, {\bf 69} 306 (1992)
\bibitem[Yee (1966)]{Yee}
Yee, K.S., ``Numerical Solution of Initial Boundary Value Problems Involving
Maxwell's Equations in Isotropic Media,'' {\it IEEE Trans. Ant. Prop.}, 
{\bf AP-14} (1966) 302

\end{thebibliography}
\end{document}